\documentclass{ifacconf}

\usepackage{amsmath,amssymb,amsfonts}
\usepackage{natbib}
\usepackage{graphicx}
\usepackage{xcolor}
\usepackage{float}
\usepackage[symbol]{footmisc}
\usepackage[ruled,vlined,algo2e]{algorithm2e}

\renewcommand{\thefootnote}{\fnsymbol{footnote}}
\DeclareMathOperator*{\argmin}{arg\,min}

\newtheorem{theorem}{Theorem}[section]
\newtheorem{lemma}[theorem]{Lemma}

\newtheorem{proposition}[theorem]{Proposition}

\newtheorem{remark}[theorem]{Remark}

\begin{document}
\begin{frontmatter}
\title{Adaptive Contraction-based Control of Uncertain Nonlinear Processes using Neural Networks}  

\author{Lai Wei,} 
\author{Ryan McCloy,} 
\author{Jie Bao\thanksref{footnoteinfo}}

\thanks[footnoteinfo]{Corresponding author: Jie Bao. This work was supported by ARC Discovery Grant DP210101978.}

\address{School of Chemical Engineering, The University of New South Wales, NSW 2052, Australia (lai.wei1@unsw.edu.au, r.mccloy@unsw.edu.au, j.bao@unsw.edu.au)}

\begin{abstract}
Driven by the flexible manufacturing trend in the process control industry and the uncertain nature of chemical process models, this article aims to achieve offset-free tracking for a family of uncertain nonlinear systems (e.g., using process models with parametric uncertainties) with adaptable performance. The proposed adaptive control approach incorporates into the control loop an adaptive neural network embedded contraction-based controller (to ensure convergence to time-varying references) and an online parameter identification module coupled with reference generation (to ensure modelled parameters converge those of the physical system). The integrated learning and control approach involves training a state and parameter dependent neural network to learn a contraction metric parameterized by the uncertain parameter and a differential feedback gain. This neural network is then embedded in an adaptive contraction-based control law which is updated by parameter estimates online. 
As uncertain parameter estimates converge to the corresponding physical values, offset-free tracking, simultaneously with improved convergence rates, can be achieved, resulting in a flexible, efficient and less conservative approach to the reference tracking control of uncertain nonlinear processes. An illustrative example is included to demonstrate the overall approach. An illustrative example is included to demonstrate the overall approach.
\end{abstract}


\end{frontmatter}
\section{Introduction}
Many modern chemical processes are inherently nonlinear and typically complex -- often comprised of a number of process units, connected through material recycle, heat integration and mass flows. Chemical processes are traditionally operated around certain equilibriums, where linear controllers are designed to control the process at these setpoints, based on linearised models. Recent years have seen the challenges for modern chemical process operations, including: the time-varying market demand for products of different specifications; variation in the cost and availability of utilities (for example, power); and significant fluctuations in the specifications of raw materials. As such, modern chemical processes require flexible operation to permit manufacturing end products with different specifications and time-varying production rates to meet the market demand and reduce operational costs. To promptly respond to supply chain fluctuations, operation practices of modern chemical processes are trending towards the flexible paradigm of smart plants (see, e.g.,~\citealp{zhang2016optimal,chokshi2008drpc}). 

Moreover, for many physical processes, obtaining process models from first principles are often accompanied by parametric uncertainty or parameter modelling error. Incorrect parameter modelling can result in significant performance loss, which from the perspective of manufacturing processes, can lead to end products of substandard and non-salable grades. Even so, sufficient understanding of the process (e.g., via guidance from the process unit manufacturer) can result in physically reasonable bounds for the system model, inside which the system parameters can be expected to vary from unit to unit (see, e.g., the variable nature of heat transfer in heat-exchanger units in   \citealp{varga1995controllability}). 
As a consequence, an adaptive control strategy to track time-varying target trajectories, to remain flexible to market demand, is required that is also capable of providing certificates of stability in the presence of parameter modelling error, as required for efficient and safe process control. 

One particularly attractive approach for time-varying reference tracking of nonlinear systems involves the contraction theory \citep{lohmiller1998contraction} framework. Following this formulation, a system's incremental stability properties can be assessed (e.g., between the plant state and target trajectory) through the study of differential dynamics. Analysis and controller synthesis simultaneously involves finding a control contraction metric (CCM) and controller pair that describe the contraction properties of the closed-loop system~\citep{manchester2017control}. Relative to popular Lyapunov based approaches, contraction theory offers the advantage of reference-flexible (i.e., reference-independent) analysis and control design (i.e., the control structure doesn't require redesign as the reference changes). Whilst contraction-based designs do offer some inherent robustness properties, metric and controller synthesis can be challenging for nonlinear systems subject to parametric uncertainty. 

An increasingly popular approach for uncertainty handling is to exploit the universal approximation characteristics of neural networks for both system identification and control (see e.g.,~\citealp{dai2013dynamic,chen1990non}). Machine learning techniques can be additionally utilised to provide learning of unmodelled or incorrectly modelled parameters~\citep{lee2018machine}, with the advantage that any offline trained neural networks may be used for further online training or tuning, providing potential for an additional level of flexibility in the controller. This philosophy is naturally befitting for the identification and control of uncertain nonlinear processes; however, it is not feasible to simply operate a chemical processes using random operating conditions to generate process data to learn an accurate model and controller due to stability/safety concerns. One approach is to train a neural network using the process model and refine it using real-world online plant data \citep{shin2019reinforcement}.

To exploit the uncertainty handling properties of neural networks and overcome the safe online learning obstacles of neural network embedded control, recent developments~\citep{wei2021control,wei2021discrete} have resulted in a discrete-time contraction control framework which facilitates safe and robust neural network training.~\cite{wei2021control} use a model with
uncertain parameters (which characterizes the inherent un-
certain nonlinear nature of modern processes) to generate training 
data, which can be done safely offline for an explicit range
of uncertainty in the system model. Contraction-based
analysis is then performed for the full range of system uncertainty
to ensure the contraction-based controller to be robust. In this
way, provided the actual system model behaves inside the
family of models considered, efficient and stabilizing control
combined with online parameter learning can be achieved.

Inspired by the philosophy proposed by \cite{shin2019reinforcement}, we propose in this article an efficient and flexible control approach method to ensure offset-free tracking with tailored rates of convergence of uncertain nonlinear systems, by leveraging our previous works~\citep{wei2021control,wei2021discrete}. An adaptive neural network-embedded contraction-based controller is first developed, with stability certificates (in terms of boundedness), to ensure bounded tracking of discrete-time nonlinear systems with parametric uncertainty. Then, an online parameter identification algorithm, is incorporated to ensure the system model tends towards the physical system description, leading to correct reference trajectory generation and hence offset-free tracking in the presence of system parameter and system target variation. Due to the adaptive nature of the neural network embedded contraction-based controller, as the online parameter estimates converge to the physical system values, the tracking performance is proportionally improved (in terms of convergence rate). Moreover, since the proposed integrated approach permits a parameter-dependent controller and contraction metric pair, conservatism and computational difficulties associated with finding a common metric and control law, for the full range of parameter uncertainty, are significantly reduced.

The remainder of this article is structured as follows. The prerequisite contraction theory tools are presented in Section \ref{sec:pre}, followed by description of the problem and overall approach in Section \ref{sec:pro}. Section \ref{sec:ctr} presents the development of an adaptive neural network embedded contraction controller with online parameter identification and controller updates. Section \ref{sec:exa} demonstrate the method via a numerical example and Section \ref{sec:conclusion} concludes this article. 

\textbf{Notation.} Denote by $f_k = f(x_k)$ for any function $f$. $B(\cdot,\cdot)$ represents a ball area centred at the first argument with second argument as the radius. The leading principle minor of $M_{NN}$ is defined as $|M_{NN_{(1,i)}}|$ and similarly for $\Omega_{(1,i)}$.

\section{Preliminaries} \label{sec:pre}
We first consider a discrete-time nonlinear control affine system without uncertainty:
\begin{equation}\label{equ:pre cer sys}
    x_{k+1} = f(x_k) + g(x_k)u_k,
\end{equation}
where state and control are $x_k \in \mathcal{X} \subseteq \mathbb{R}^n$ and $u_k \in \mathcal{U} \subseteq \mathbb{R}^m$. The corresponding differential system of \eqref{equ:pre cer sys} is as follows:
\begin{figure}
    \begin{center}
        \includegraphics[width=0.8\linewidth]{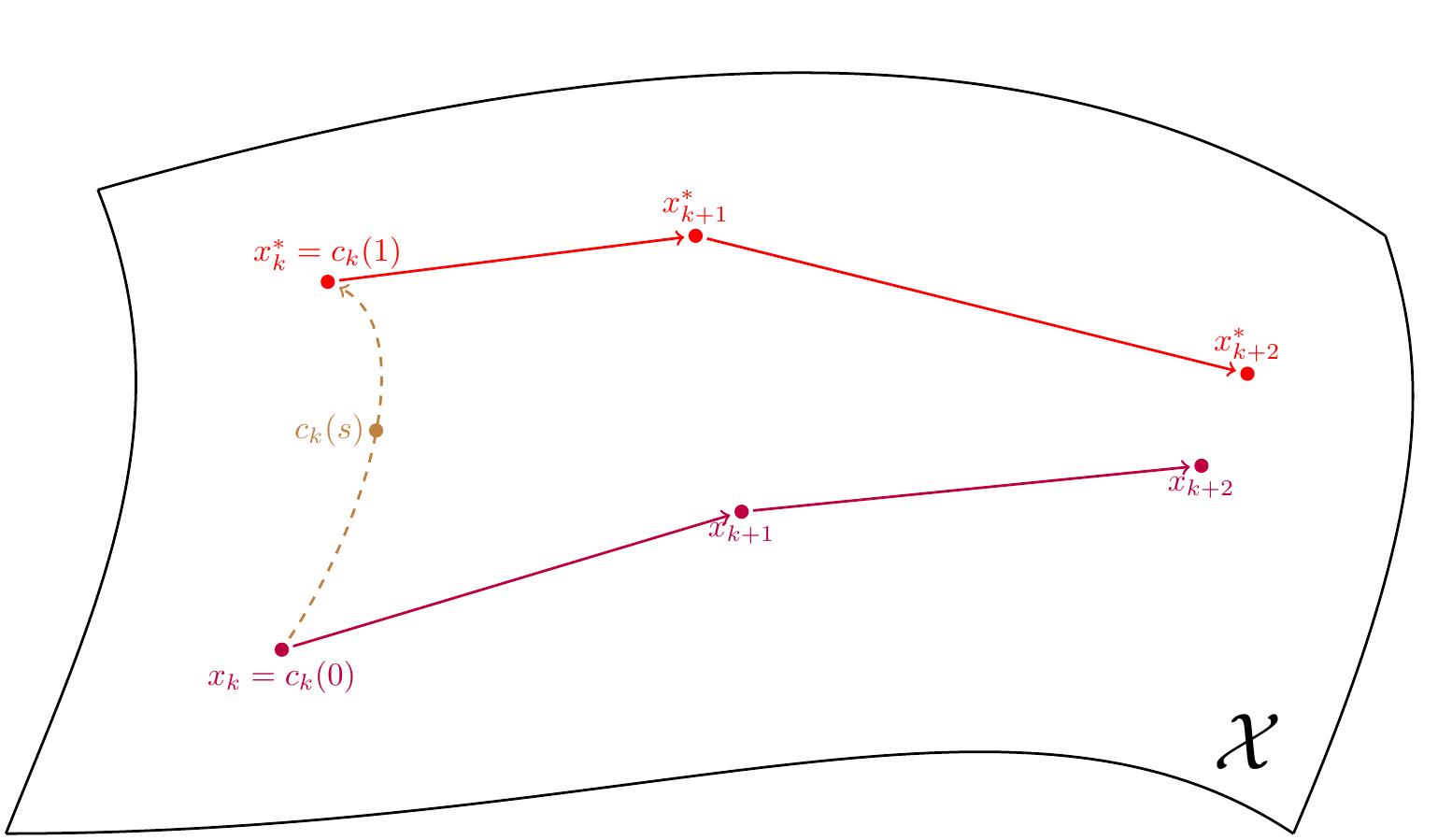}
        \caption{Path parameterised by parameter $s$.}
        \label{fig:rie_geo}
    \end{center}
\end{figure}
\begin{equation}\label{equ:pre cer dif sys}
    \delta_{x_{k+1}} = A(x_k)\delta_{x_k} + B(x_k)\delta_{u_k},
\end{equation}
where Jacobian matrices of $f$ and $g$ in \eqref{equ:pre cer sys} are defined as $A:=\frac{\partial (f(x_k) + g(x_k)u_k)}{\partial x_k}$ and $B:=\frac{\partial (f(x_k) + g(x_k)u_k)}{\partial u_k}$  respectively, $\delta_{u_k} := \frac{\partial u_k}{\partial s}$ and $\delta_{x_k}:=\frac{\partial x_k}{\partial s}$ are vectors in the tangent space $T_x\mathcal{U}$ at $u_k$ and tangent space $T_x\mathcal{X}$ at $x_k$ respectively, where $s$ parameterises a path, $c(s): [0,1] \rightarrow \mathcal{X}$ between two points such that $c(0) = x, c(1) = x^*  \in \mathcal{X}$ (see Fig. \ref{fig:rie_geo}). If we consider a state-feedback control law for the differential dynamics \eqref{equ:pre cer dif sys}, i.e.,
\begin{equation}\label{equ:pre dif fed}
    \delta_{u_k} = K(x_k) \delta_{x_k},
\end{equation}
where function $K$ is a state dependent function. Then, Theorem \ref{thm:pre ctr con} describes the contraction condition for a discrete-time nonlinear control affine system in \eqref{equ:pre cer sys} as follows,
\begin{theorem}[\citealp{wei2021discrete}]
    \label{thm:pre ctr con}
    For a discrete-time nonlinear system \eqref{equ:pre cer sys}, with differential dynamics \eqref{equ:pre cer dif sys} and differential state-feedback controller \eqref{equ:pre dif fed}, provided a uniformly bounded DCCM, $M(x_k)$, satisfying, 
    \begin{equation}\label{equ:pre ctr con} 
        (A_k+B_kK_k)^\top M_{k+1}(A_k+B_kK_k) - (1-\beta)M_{k} < 0,
    \end{equation}
    exists, then the closed-loop system is contracting for some constant $0 < \beta \leq 1$. Furthermore, the closed-loop system is incrementally exponentially stable,  i.e.,
    \begin{equation}
        \label{equ:pre exp sta}
        |x_k-x^*_k| \leq  R e^{-\lambda k} |x_0 - x^*_0|,
    \end{equation}    
    for some constant $R$, convergence rate $\lambda$ and any feasible reference trajectory $(\hat{x}^*, \hat{u}^*)$, satisfying \eqref{equ:pre cer sys}, where $x_k$ is the state value at time-step $k$ with the initial condition $x_0$.
\end{theorem}
A region of state space is called a contraction region if condition \eqref{equ:pre ctr con} holds for all points in that region. In Theorem \ref{thm:pre ctr con}, $M$ is a metric used in describing the geometry of Riemannian space, which we briefly present here. We define the Riemannian distance, $d(x,x^*)$, as (see, e.g., \citealp{carmo1992})
\begin{equation}\label{equ:Riemannian distance and energy}
    \begin{aligned}
    d(x,x^*) = d(c) :=\int_0^1 \sqrt{\delta^\top_{c(s)}M(c(s))\delta_{c(s)}}ds,
    \end{aligned}
\end{equation}
where $\delta_{c(s)} := \frac{\partial c(s)}{\partial s}$.
The shortest path in Riemannian space, or \textit{geodesic}, between $x$ and $x^*$ is defined as 
\begin{equation}\label{equ:geodesic}
    \gamma(s) :=\argmin_{c(s)} {d(x,x^*)}.
\end{equation}

Leveraging Riemannian tools, one feasible feedback tracking controller for  \eqref{equ:pre cer sys}, can be obtained by integrating the differential feedback law \eqref{equ:pre dif fed} along the geodesic, $\gamma(s)$ \eqref{equ:geodesic}, as
\begin{equation}\label{equ:pre ctl int}
    u_k = u^*_k + \int_0^1K(\gamma(s))\frac{\partial \gamma(s)}{\partial s}\,ds.
\end{equation}
\section{Problem Formulation and Approach} \label{sec:pro}
    \subsection{System Description}
    Herein, we consider the following discrete-time control affine nonlinear system with parametric uncertainty (for further discussion see \citealp{xu2011survey,barmish1985necessary}):
    \begin{equation}\label{equ:pro unc sys}
        x_{k+1} = f(r,x_k) + g(r,x_k)u_k,
    \end{equation}
    where functions $f$ and $g$ are smooth along the $x$ direction and Lipschitz continuous along $r$ and $k$. The additional argument vector (cf. \eqref{equ:pre cer sys}), $r$, denotes the bounded uncertain system parameters given by,
    \begin{equation}\label{equ:pro unc bnd}
        r \in \mathcal{R} = \{r \in \mathbb{R}^\ell ~|~ r_{min}^i \leq r^i \leq r_{max}^i ~|~ i=1,\dots,\ell\},
    \end{equation}
    where $r^i$ is the $i$-th element of $r$, $r_{min}^i$ and $r_{max}^i$ are the lower and upper bound of the $i$-th element. 
    The corresponding differential dynamics at a certain parameter value $r$ can be determined, i.e.,
    \begin{equation}\label{equ:pro unc dif sys}
        \delta_{x_{k+1}} = A(r,x_k)\delta_{x_k} + B(r,x_k)\delta_{u_k},
    \end{equation}
    where $A(r,x_k):=\frac{\partial (f(r,x_k) + g(r,x_k)u_k)}{\partial x_k}$ and $B(r,x_k):=\frac{\partial (f(r,x_k) + g(r,x_k)u_k)}{\partial u_k}$ are Jacobian matrices evaluated at $x_k$ and $u_k$, given $r$. 
    
    \subsection{Objective and Approach}\label{sec:obj app}
     Time-varying state and control targets $(x^*_{k+1},x^*_k,\hat{u}^*_k)$  for \eqref{equ:pro unc sys} are generated using an estimate of the uncertain parameter $\hat{r}_k$, satisfying
    \begin{equation}\label{equ:pro unc ref}
        x_{k+1}^* = f(\hat{r}_k,x_k^*) + g(\hat{r}_k,x_k^*)\hat{u}_k^*.
    \end{equation}
    These solutions, $(x^*_k,x^*_{k+1},\hat{u}^*_k)$, are only feasible solutions for the actual system dynamics \eqref{equ:pro unc sys} when the estimated parameter matches the physical system value, i.e.,  $\hat{r}_k = r$. As a result, generating control references subject to parameter modelling error will result in incorrect control targets and hence state tracking offsets (see, e.g., \citealp{wei2021control}).

    Thus, the control objective is to force online convergence of the parameter estimate, $\hat{r}_k \to r$, whilst ensuring stability (to state reference targets, i.e., $x \to x^*$) for the full range of parameter variation. In addition, exploiting improved parameter modelling for improved control performance is also desired.
    
    To achieve our offset-free tracking objective, which relies on generating the correct reference and hence correct model parameters, an adaptive neural network embedded contraction controller will be trained, as a function of the system state and current parameter estimates, and employed in the closed-loop with an optimisation-based parameter identification module. The contraction-based controller will provide a stable framework, from which online learning can be conducted, which in turn is self-tuned via updates of new parameter estimates. As the online parameter identification converges to the real system value, the tracking performance of the overall control scheme is improved, by: 1) feasible state and control reference pairs can be generated, resulting in error-free convergence of the state trajectory, $x_k$ to the target trajectory $x_k^*$; and 2) adapting the metric and controller gains, resulting in state-to-reference convergence rates approaching desired (design) values.

        \section{Contraction-based Learning Control} \label{sec:ctr}
    The first step in the proposed approach is to provide guaranteed contraction for uncertain nonlinear systems, for which the DCCM is permitted to vary with time (as parameter estimates are updated), by extending the methods developed by \cite{wei2021discrete}. The second step is to provide a generalised approach to identifying the parametric uncertainty online. By ensuring uncertain parameter estimates converges to the physical value, offset-free tracking simultaneously with improved convergence rates can be achieved.
    
    \subsection{Adaptive contraction control for uncertain systems}
   Firstly, consider the function pair $(M(x_k,\hat{r}),K(x_k,\hat{r}))$ which satisfies Theorem \ref{thm:pre ctr con} for a specific value of the uncertain parameter $\hat{r} \in \mathcal{R}$, i.e., for the system described by \eqref{equ:pro unc sys}--\eqref{equ:pro unc dif sys} satisfying 
   \begin{equation}\label{equ:unc ctr con} 
        (A_k+B_k K_k)^\top M_{k+1}(A_k+B_kK_k) - (1-\beta)M_{k} < 0,
    \end{equation}
   where $A_k := A(x_k,\hat{r})$, $B_k := B(x_k,\hat{r})$, $M_k := M(x_k,\hat{r})$, $M_{k+1} := M(x_{k+1},\hat{r})$ and $K_k := K(x_k,\hat{r})$. We then consider the following robustness Lemma.
   \begin{lemma}[\citealp{wei2021discrete}]\label{lem:bounded}
   For the DCCM-based controller \eqref{equ:pre ctl int} that ensures a system without uncertainty \eqref{equ:pre cer sys} (i.e., $\hat{r} = r$) is contracting, when parametric uncertainty is present  \eqref{equ:pro unc sys} (i.e., $\hat{r} \neq r$), the state trajectory, $x$, is driven by \eqref{equ:pre ctl int} to the bounding ball around the target reference, $x^*$, as 
    \begin{equation}\label{eq:bball}
    d(\gamma_{k+1}) \leq (1-\beta)^{\frac{1}{2}} d(\gamma_k) + \sqrt{\alpha_2}\max_{x_k} \| g(r,x_k) \| \|u^*_k - \hat{u}_k^*\|.
    \end{equation}
    with at least rate $\beta$, where $u^*_k$ represents the control input reference generated using the true parameter value $r$.
   \end{lemma}

Then, extending the state-dependent DCCM and feedback gains to be additionally dependent on uncertain parameter estimates (evolving with time as parameter estimates are updated), denoted by $(\hat{M}(x_k,\hat{r}_k),\hat{K}(x_k,\hat{r}_k))$ (note that, e.g.,  $\hat{M}(x,\hat{r}_1)$ is not necessarily equal to $\hat{M}(x,\hat{r}_2)$, for $r_1,r_2 \in \mathcal{R}$, irrespective of the time-step $k$), where the state and parameter-dependent differential feedback gain $\delta_u = \hat{K}(x_k,\hat{r}_k)$ results in the adaptive contraction-based controller
(integrating along the geodesic, $\gamma(s)$ \eqref{equ:geodesic})
\begin{equation}\label{equ:adapt ctl int}
    u_k = \hat{u}^*_k + \int_0^1\hat{K}_k(\gamma(s))\frac{\partial \gamma(s)}{\partial s}\,ds,
\end{equation}
where $\hat{u}^*_k$ is the current control reference satisfying \eqref{equ:pro unc ref} for the desired state $x_k^*$ and current parameter estimate $\hat{r}_k$.
We then propose the following performance results.

\begin{proposition}\label{prop:unc adapt}
Consider the uncertain system \eqref{equ:pro unc sys} with differential dynamics \eqref{equ:pro unc dif sys}, and adaptive  pair \sloppy$(\hat{M}(x_k,\hat{r}_k),\hat{K}(x_k,\hat{r}_k))$, satisfying a contraction condition \eqref{equ:unc ctr con} for each instance of $\hat{r} \in \mathcal{R}$ \eqref{equ:pro unc bnd} with desired contraction rate $\beta$. Provided this pair additionally satisfies
 \begin{equation}\label{equ:slower broader}
        \begin{aligned}
            \left(A_k(r)+B_k(r)\hat{K}_k(r)\right)^\top \hat{M}_{k+1}(r)\left(A_k(r)+B_k(r)\hat{K}_k(r)\right) \\
            - (1-\beta_\ell)\hat{M}_{k}(r) < 0. \qquad \forall r \in \mathcal{R},
        \end{aligned}
\end{equation}
the adaptive controller \eqref{equ:adapt ctl int} ensures bounded convergence to state references, with a contraction rate of at least $\beta_\ell$. Moreover, as the parameter estimate, $\hat{r}$ converges to the true system value $r$: i) offset-free tracking is achieved, and; ii) the contraction rate approaches the desired rate $\beta$. 
\end{proposition}
\begin{pf}
Since each parameter estimate $\hat{r}$ is coupled with a metric $\hat{M}(x_k,\hat{r})$ and differential feedback gain $\hat{K}(x_k,\hat{r})$ satisfying \eqref{equ:unc ctr con} (and hence \eqref{equ:pre ctr con}), under Theorem \ref{thm:pre ctr con} and Lemma \ref{lem:bounded}, the metric $\hat{M}(x_k,\hat{r}_k)$ is a DCCM for \eqref{equ:pro unc sys}, which achieves bounded tracking when driven by the controller \eqref{equ:adapt ctl int}. Hence, as $\hat{r} \to r$, $\hat{u}^* \to u^*$ and $d_{\gamma} \to 0$, yielding i).
    
Consider the DCCM and feedback gain for the true parameter $r$ as $(M,K)$ and $(\hat{M},\hat{K})$ for on particular parameter estimate $\hat{r}$. Provided a slower contraction rate $0< \beta_\ell \leq \beta$ exists such that \eqref{equ:slower broader} holds, 
where $\beta_\ell$ is a Lipschitz function of $\hat{r}$ such that $ \beta \geq \beta_\ell \geq \beta - L\|\hat{r} - r\| > 0 $, then, as $\hat{r}_k \to r$, $(\hat{M},\hat{K}) \to (M,K)$, we have $\beta_\ell \to \beta$, and the result in ii). 
\end{pf}

    
\begin{remark}
    To see that satisfying \eqref{equ:slower broader} is reasonable, Let $\mathcal{S}(x,\beta) = \{x\in \mathcal{X} | \delta_{x_{k+1}}^\top M_{k+1} \delta_{x_{k+1}} - (1-\beta) \delta_{x_k}^\top M_k \delta_{x_k}^\top < 0 \}$ denote a contraction region. Suppose we then have a desired rate $\beta_d$ and slower rate $\beta_s \leq \beta_d$ (ensured by the existence of $\beta_d$), then, $S_s(x,\beta_r)$ denotes the larger contraction region where $\mathcal{S}_d(x,\beta_d) \subseteq \mathcal{S}_s(x,\beta_s)$. As such, robustness to parametric uncertainty can be achieved based on the assumption that the contraction rate can always be reduced to expand the region of contraction and account for small modelling discrepancies (still resulting in bounded convergence), provided one instance of the contraction condition in \eqref{equ:unc ctr con} can be satisfied.
\end{remark}
    
    The advantage of Proposition \ref{prop:unc adapt} lies in the reduction of conservatism and computational burden of solving \eqref{equ:unc ctr con} for a common metric, i.e., finding one pair $(M,K)$ satisfying \eqref{equ:pre ctr con} $\forall r\in \mathcal{R}$, permitting improved performance (i.e., faster / less conservative convergence rates) and amenability to practical system models (by considering existence and computational difficulty of finding a common metric). Moreover, this framework permits the extension to parameter varying convergence rates, i.e., modifying \eqref{equ:unc ctr con} to include $\beta(r)$.
    
    \subsubsection{Neural network-based DCCM Control}
\begin{figure}[t]
        \begin{center}
            \includegraphics[width=0.8\linewidth]{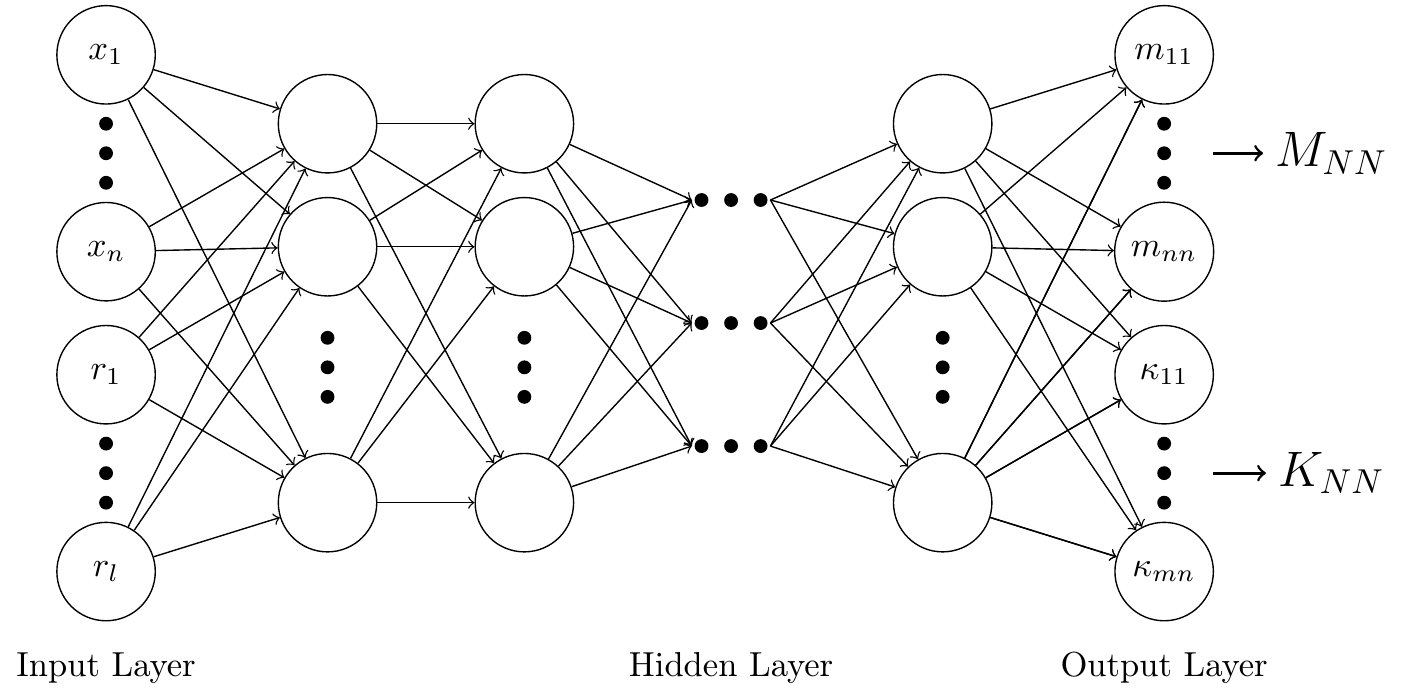}
            \caption{Neural network represented DCCM and differential controller.}
            \label{fig:nn dccm k}
        \end{center}
    \end{figure}

    To learn each pair $(\hat{M},\hat{K})$ satisfying \eqref{equ:unc ctr con} given each parameter value $\hat{r} = r \in \mathcal{R}$, a neural network is used. As such, a neural network, as shown in Fig. \ref{fig:nn dccm k}, is used to represent the adaptive function pair, $(\hat{M},\hat{K})$, satisfying the contraction condition \eqref{equ:unc ctr con}. The inputs of the neural network are the system states and uncertain parameter values, and the outputs of the neural network are entries of matrix functions $(M_{NN},K_{NN})$, i.e., the neural network represented $(\hat{M},\hat{K})$. 
    
\begin{figure}[t]
        \begin{center}
            \includegraphics[width=\linewidth]{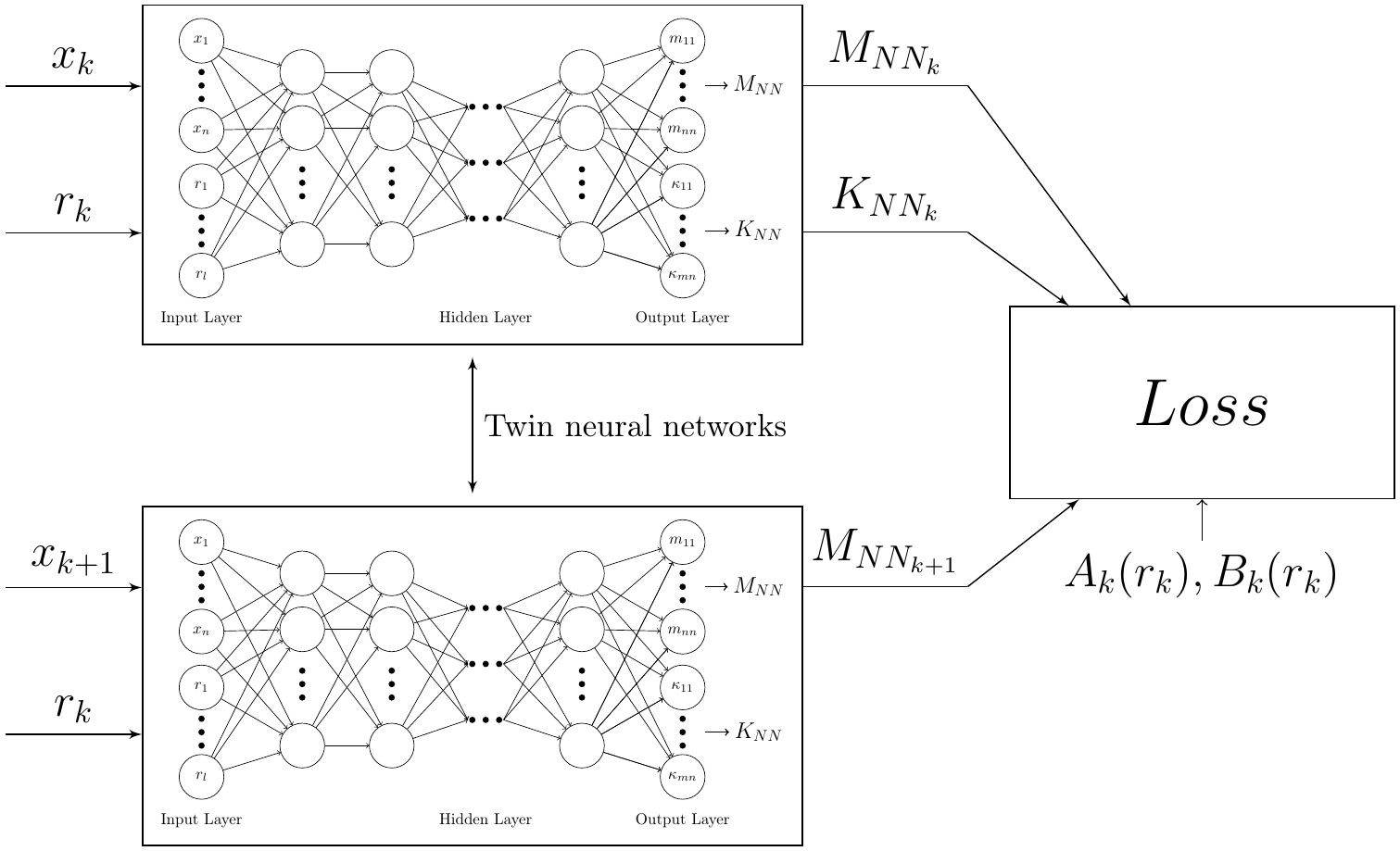}
            \caption{Training process block diagram.}
            \label{fig:siam}
        \end{center}
    \end{figure}

    The neural network requires a targeted loss function, $L$, for effective training, and is defined as
    \begin{equation}\label{equ:ctr lean func}
        \begin{aligned}
        L_{M_i} =&
        \begin{cases}
            -(|M_{NN_{(1,i)}}| - \epsilon_i)  &if ~(|M_{NN_{(1,i)}}| - \epsilon_i) \leq 0\\
            0 \ &else
        \end{cases}\\
        L_{\Omega j} =& 
        \begin{cases}
            -(|\Omega_{(1,j)}| - \epsilon_j) \quad \  &if ~(|\Omega_{(1,j)}| - \epsilon_j) \leq 0\\
            0 \ &else
        \end{cases}\\
        L =& \sum_i L_{Mi} + \sum_j L_{\Omega j}, \qquad \text{where} \\
        \Omega :=& -(A_k(r)+B_k(r)K_{NN_k}(r))^\top M_{{NN}_{k+1}}(r)(A_k(r) \dots \\
        &+B_k(r)K_{NN_k}(r)) + (1-\beta)M_{NN_k}(r)
        \end{aligned}
    \end{equation}
    and $\epsilon_i$ and $\epsilon_i$ are some small positive values. This loss function represents the contraction condition \eqref{equ:unc ctr con}, ensuring that it holds for all possible two-step trajectories $\{x_k,x_{k+1}\}$ and uncertain parameter values $r$. Since the contraction condition \eqref{equ:unc ctr con} includes both current and future time-steps, the usage of a Siamese network structure is well suited, as shown in Fig. \ref{fig:siam}. 
    
    Finally, Algorithm \ref{algo:training} (adapted from \citealp{wei2021discrete}) is used to generate a data set based on a model of the system and train the neural network. By computing a feasible function pair $(M_{NN},K_{NN})$, we can implement the following adaptive controller, based on the neural network learned DCCM and differential feedback gain, 
    \begin{equation}\label{equ:ctr nn con ctl int}
        u_k = u^*_k + \int_0^1K_{NN}(\gamma_{NN}(s))\frac{\partial \gamma_{NN}(s)}{\partial s}\,ds,
    \end{equation}
    where $\gamma_{NN}$ is the geodesic, connecting the current state $x_k$ and the reference state $x_k^*$, calculated using the DCCM $M_{NN}$.
    
    \begin{algorithm2e}
        \SetAlgoLined
        \For{$u_k \in \mathcal{U}$}{
            \For{$x_k \in \mathcal{X}$}{
                \For{$r \in \mathcal{R}$}{
                    Calculate $x_{k+1}$ using \eqref{equ:pro unc sys}.\\
                    Compute Jacobian matrices $A_k,B_k$.\\
                    Store $\{r,x_k,x_{k+1},A_k,B_k\}_i$ in data set $\mathcal{D}$.
                }
            }
        }
        All elements, $\{r,x_k,x_{k+1},A_k,B_k\}_i \in \mathcal{D}$, form a batch.\\
        \For{$iteration \leq upper~ limit$}{
            \For{each $\{r,x_k,x_{k+1},A_k,B_k\}_i \in \mathcal{D}$}{
            Feed $x_k$ and $r$ into the first neural network.\\
            Construct $M_{NN_k}$ and $K_{NN_k}$.\\
            Feed $x_{k+1}$,$r$ into the second neural network.\\
            Construct $M_{NN_{k+1}}$.\\
            Calculate the loss, $L_i$, for the $i$-th element .\\
            }
        Calculate the total loss $L_t = \sum_i{L_i}$ for a batch.\\
        Proceed with backpropagation.\\ 
        \If{$L_t<tolerance$}{Break.}
        }
        Save the neural network represented $M_{NN}$ and $K_{NN}$.
        \caption{Data Generation and Training}\label{algo:training}
    \end{algorithm2e}

    \subsubsection{Analysis}
    Firstly, we will present a result from \cite{wei2021discrete} that describes the existence of a contraction rate with respect to the learned pair $(M_{NN},K_{NN})$.
    \begin{theorem}[\citealp{wei2021discrete}]\label{cly:lea xur}
     The uncertain nonlinear system \eqref{equ:pro unc sys}, with fixed parametric uncertainty, $r$ \eqref{equ:pro unc bnd}, differential dynamics \eqref{equ:pro unc dif sys} and neural network embedded contraction based controller \eqref{equ:ctr nn con ctl int}, is locally contracting (for each instance of $r\in \mathcal{R}$) within a bound about the reference, $x^*$, provided Algorithm \ref{algo:training} finds locally contracting balls, $B_{x,i}$, centered at $x_k$ for each $i$-th data element, forming the area of interest, $\mathcal{X}_r \subseteq \underset{i}{\bigcup} B_{x,i}$, as a contraction region with minimum convergence rate, $\lambda_{\mathcal{S}_x,min}$, given by,
       \begin{equation}
           \lambda_{\mathcal{S}_x,min} = \min_i (\lambda - L_{xur}||\xi_i||),
       \end{equation}
       where $L_{xur}$ is a Lipschitz constant.
    \end{theorem}

    Theorem \ref{cly:lea xur} presents a way to implement a numerical DCCM learning algorithm for a large contraction region (e.g., an area of interest) by studying a relatively smaller region of balls, under fixed parametric uncertainty. Consequently, this permits the search for a pair $(M,K)$ at a particular value of $r$ to be transformed into a neural network training problem using discrete (meshed) training data.

    Hence, from Proposition \ref{prop:unc adapt} and Theorem \ref{cly:lea xur}, contraction of the system state to the reference trajectory within a bound is guaranteed for nonlinear systems centred at specific values of parametric uncertainty. 
    However, to achieve offset-free tracking, (as explained in Section \ref{sec:obj app}) the reference generator requires the estimate for the uncertain parameter to match the physical system value. This will be addressed in the following section.
    \subsection{Online Parameter Identification}
    As concluded in the previous section, the combination of Proposition \ref{prop:unc adapt} and Theorem \ref{cly:lea xur} provides a stable (in terms of boundedness) framework from which to employ online parameter identification methods (e.g. moving horizon estimation, Kalman filter, neural network estimation) and hence online adaptive tuning of the neural network embedded controller. Here we will employ the efficient and well studied moving horizon estimation approach (see, e.g., \citealp{rawlings2017model,olivier2017performance} for further details),  which involves recursively solving the following least-squares optimisation problem. 
\begin{equation}\label{eq:MHEP}
\begin{aligned}
        & \qquad \quad \min_{\mathbf{\hat{r}}}\sum_{i=k-N}^{k} \| \hat{x}_{i} - {x}_{i} \| \\
        s.t. ~~~~ &\hat{x}_{i} = f(\hat{r}_{i-1},x_{i-1}) + g(\hat{r}_{i-1},x_{i-1})u_{i-1},\\
        &\qquad \quad \,\,\,\, r_{min} \leq \hat{r}_i \leq r_{max},
\end{aligned}
\end{equation}
where $\mathbf{\hat{r}} = (\hat{r}_{k-N},\cdots,\hat{r}_{k-1})$ is a sequence of parameter estimates, $\hat{x_i}$ is the estimated state at time step $i$ using $\hat{r}$ and \eqref{equ:pro unc sys}, and $N$ is the estimation horizon.  

An updated parameter estimate, $\hat{r}_k$, is provided by solving \eqref{eq:MHEP} for the time interval $[k,k+1)$, whereby the current step estimate is assigned the most recent estimate value (last element in $\mathbf{\hat{r}}$), i.e., $\hat{r}_k = \hat{r}_{k-1}$. This is reasonable under the initial assumption that the uncertain parameter is a constant. 
Moreover, weightings for each step in the horizon, and for each element in $r$, can be added to emphasise the relative significance of the parametric  identification error. The length of the estimation horizon, $N$, is chosen by considering the identifiability of the uncertain parameter, $r$ (see \citealp{olivier2017performance}), as determined by the parameter's dimensionality and the sufficiency of the data generated in Algorithm \ref{algo:training}. Note that the estimated value, $\hat{r}_k$, is forced to lie inside the known uncertain bound $(r_{min},r_{max})$, satisfying \eqref{equ:pro unc bnd}, and converges to the physical value, $r$, as $\hat{x}$ converges to $x$, leading to the following proposition for the overall closed-loop system.  
    \begin{proposition}\label{propo}
        The discrete-time nonlinear system \eqref{equ:pro unc sys} with parametric uncertainty, $r$ \eqref{equ:pro unc bnd}, differential dynamics \eqref{equ:pro unc dif sys}, adaptive neural network embedded contraction-based controller \eqref{equ:ctr nn con ctl int}, and online parameter estimator (via solution to~\eqref{eq:MHEP}) achieves offset-free tracking (i.e., the Riemannian distance shrinks to a bound with some contraction rate and $\lim_{k \rightarrow \infty}|x_k -x_k^*| = 0$ as the parameter estimation converges without error to the physical value) if the pair $(M_{NN},K_{NN})$ is found offline satisfying Theorem \ref{cly:lea xur} and Proposition \ref{prop:unc adapt}, and solution to \eqref{eq:MHEP} converges online. Moreover, as the parameter estimate converges, the rate of convergence approaches the desired rate.  
    \end{proposition}   
    \begin{pf}
        If the estimation error $\lim_{k \rightarrow \infty} |\hat{r}-r| = 0$, then the reference generated by \eqref{equ:pro unc ref} is a feasible trajectory of \eqref{equ:pro unc sys}. Since the adpative controller \eqref{equ:ctr nn con ctl int} uses the function pair $(M_{NN},K_{NN})$ solved from Algorithm \ref{algo:training}, then, from Theorem \ref{thm:pre ctr con} and Proposition \ref{prop:unc adapt}, we have that the system is contracting and without error with a rate approaching the desired value.  
    \end{pf}

    \textbf{In summary}, 
    process history is sent to an online parameter identification module to estimate the uncertain parameter, $\hat{r}_k$ via solution to an online optimisation problem~\eqref{eq:MHEP}. The estimated parameter $\hat{r}_k$ is sent to the reference generator, which computes the correct state and control references when the parameter is correctly identified, i.e., $\hat{r}_k = r$ (See Section \ref{sec:obj app}). To compute the control law, the geodesic \eqref{equ:geodesic}, connecting the current state, $x_k$, and target state, $x^*_k$, is first calculated numerically using the neural network represented metric, $M_{NN}$, and current parameter estimate, $\hat{r}_k$. Then, the control input \eqref{equ:ctr nn con ctl int} is calculated using this geodesic, the current parameter estimate and the neural network represented feedback gain, $K_{NN}$ (obtained via Algorithm \ref{algo:training}). Under Proposition \ref{propo}, the proposed closed-loop scheme ensures offset-free tracking with desired convergence rates for nonlinear systems subject to parametric uncertainty of the form in \eqref{equ:pro unc sys}.

\section{Example} \label{sec:exa}
Consider the following nonlinear continuously stirred tank reactor (CSTR) process~\citep{mccloy2021differential}, discretised with time-steps of $0.1$ hour, 
\begin{equation}\label{equ:sim cstr certain}
     \begin{bmatrix}
     x_{1_{k+1}}\\ 
     x_{2_{k+1}}
     \end{bmatrix}= 
    \begin{bmatrix}
        \begin{aligned}
            &0.9x_{1_k} + 0.1\phi_1(x_{1_k}) e^{\frac{\alpha x_{2_k}}{\alpha+x_{2_k}}} + 0.1(1-\zeta)x_{1_k}\\
            &0.9x_{2_k} + 0.1\phi_2(x_{1_k}) e^{\frac{\alpha x_{2_k}}{\alpha+x_{2_k}}} + u_k
        \end{aligned}
    \end{bmatrix},
\end{equation}where $x_{1_k}$, $x_{2_k}$ and $u_k$ denote the normalised reactant concentration, reactor temperature and jacket temperature respectively, $\alpha = 0.8$, $\zeta = 0.1$,   $\phi_i(x_{1_k}) = Da_i(1-x_{1_k})$, where the uncertain parameters are in the range of  $\hat{Da}_1 = [1.15,3.125]$ and $\hat{Da}_2 = [1.275,3.438]$, with ``true'' values of $Da_1 = 1.25$ and $Da_2 = 1.375$. The state and input constraints are $x_{1_k} \in [0.1,1.1]$, $x_{2_k} \in [-0.1,1.1]$ and $u_k \in [-1,1]$, respectively.  Offline neural network training to obtain the DCCM, $M_{NN}$, and feedback gain, $K_{NN}$ with respect to the desired contraction rate $\beta=0.26$, using the Siamese neural network, was completed via Algorithm \ref{algo:training} with a learning rate of $0.001$, 3 hidden layers and 15 nodes per layer, as shown in Fig. \ref{fig:nn dccm k}. A relaxed (slower) rate, $\beta_\ell = 0.21$, was also found, satisfying Proposition \ref{prop:unc adapt}. We note here, that for this CSTR problem, the proposed control approach offers increased synthesis flexibility (e.g., relative to a similar problem and approach in \citealp{wei2021control}), whereby the DCCM and feedback gain are parameterised by the uncertainty (as opposed to the computationally strict approach of searching for a common solution pair across the full range of uncertainty). 

\begin{figure}
    \begin{center}
        \includegraphics[width=\linewidth]{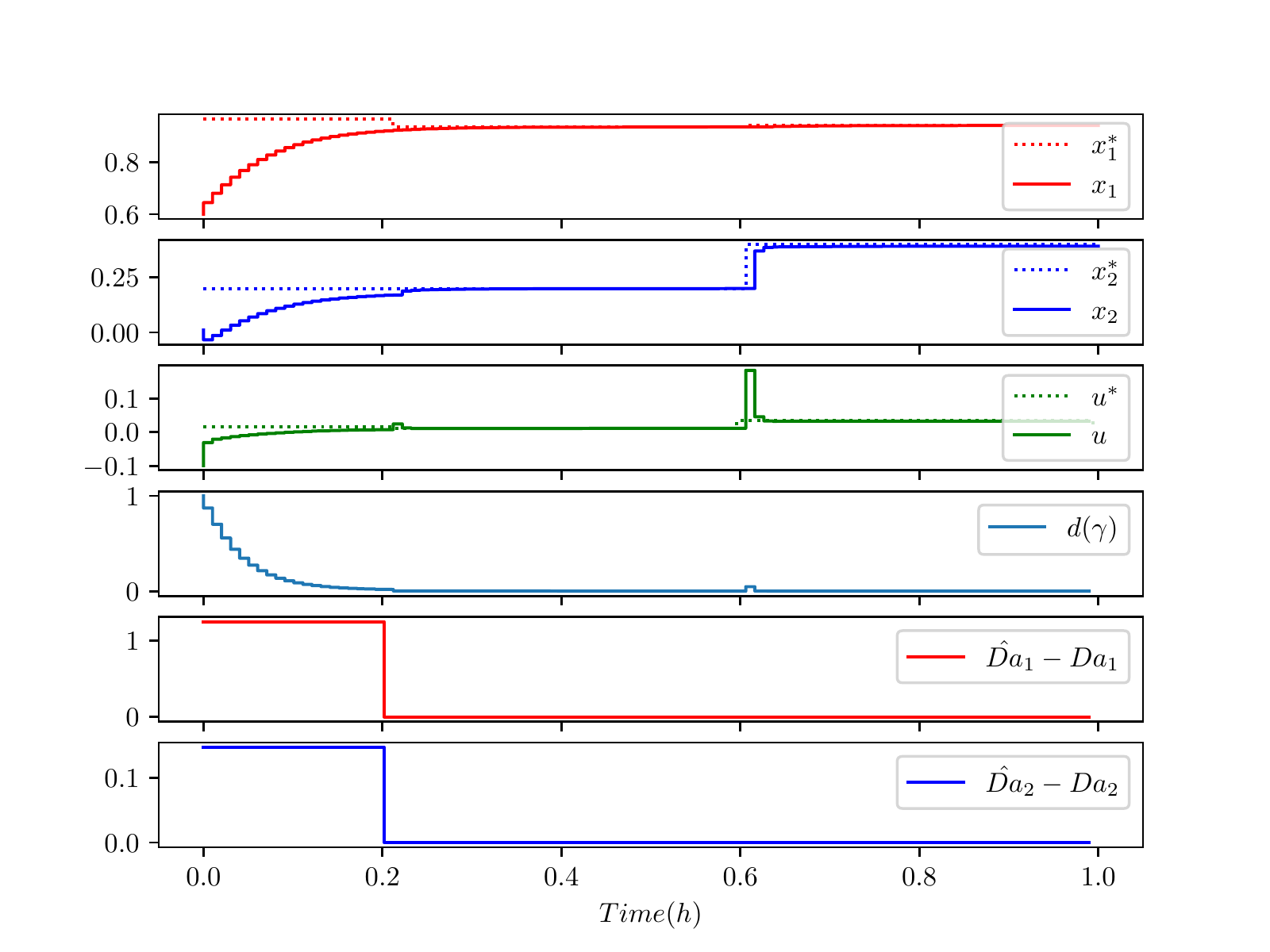}
        \caption{Simulation of a CSTR.}
        \label{fig:response}
    \end{center}
\end{figure}

The simulation result is presented in Fig. \ref{fig:response}. The initial conditions are $x_0 = [0.6,0.01]^\top$, the desired references are $x^* = [0.936,0.2]^\top$ for $[0,0.6)h$ and $x^* = [0.940,0.3]^\top$ for $[0.6,1.0]h$. Initially, $\hat{Da}_1 = 2.50$ and $\hat{Da}_2 = 1.522$ were modelled incorrectly. The online identification module was not active before time $0.2$h, and  hence incorrect references for contraction-based control were generated (see Section \ref{sec:obj app}). Therefore, the system true states $x_1$ and $x_2$ were not tracking the reference states $x_1^*$ and $x_2^*$ offset-free (although they did track within a bound, as expected under Theorem \ref{cly:lea xur} and Proposition \ref{prop:unc adapt}). The online identification module was implemented  from $0.2$h onwards (to estimate $\hat{Da_1}_k$ and $\hat{Da_2}_k$ via solving \eqref{eq:MHEP}). The estimated values of $\hat{Da}_1$ and $\hat{Da}_2$ converged to their true values. As a consequence, the system trajectory was tracking the time-varying reference without errors. The rate of convergence approached the desired rate (resulting in a $20\%$ rate improvement as $\beta_\ell \to \beta$), as per Proposition~\ref{propo}.

\section{Conclusion} \label{sec:conclusion}
In this article, an integrated identification and control approach utilising neural networks and contraction theory was proposed. An adaptive neural network-embedded contraction-based controller was developed for uncertain nonlinear systems, which guarantees bounded convergence of the system to time-varying reference targets for the full range of unknown parameter variation and without the need for controller structural redesign. The adaptive controller is updated online by new state measurements and parameter estimates. Parameter estimates are computed by an optimisation-based online identification module to facilitate correct reference generation and consequently improve the controller tracking performance with respect to tracking-error and convergence rates. The resulting closed-loop control scheme ensures offset-free tracking with desired convergence rates (when the identified model matches the physical system) for discrete-time nonlinear systems with bounded parametric uncertainty. Finally, a simulation study was presented to demonstrate the approach. 
\bibliography{ifacconf}
\appendix
\end{document}